\documentclass[12pt]{article}
\usepackage[dvips]{color}
\usepackage{epsfig}
\usepackage{amsmath}
\usepackage{graphicx}

\begin{document}
\title{\bf Thermal Equilibrium of Hagedorn and Radiation Regimes in String Gas Cosmology}
\author{J. Sadeghi$^{a,b,}$\thanks{Email:
pouriya@ipm.ir}\hspace{1mm}, H. Farahani$^{a,}$\thanks{Email:
h.farahani@umz.ac.ir}\hspace{1mm} and B.
Pourhassan$^{a,}$\thanks{Email:
b.pourhassan@umz.ac.ir}\\
$^a$ {\small {\em  Sciences Faculty, Department of Physics, Mazandaran University,}}\\
{\small {\em P .O .Box 47416-95447, Babolsar, Iran}}\\
$^b$ {\small {\em  Institute for Studies in Theoretical Physics and
Mathematics (IPM),}}\\
{\small {\em P.O.Box 19395-5531, Tehran, Iran}} } \maketitle
\begin{abstract}
In this paper, we investigate thermal equilibrium in string gas
cosmology which is dominated by closed string. We consider two
interesting regimes, Hagedorn and radiation regimes. We find that
for short strings in small radius of Hagedorn regime very large
amount of energy requested to have thermal equilibrium but for long
strings in such system a few energy is sufficient to have thermal
equilibrium. On the other hand in the large radius of Hagedorn
regime, which pressure is not negligible, we obtain a relation
between the energy and pressure in terms of cosmic time which is
satisfied by thermal equilibrium. Then we discuss about radiation
regime and find that in all cases there is
thermal equilibrium.\\\\
\noindent {\bf Keywords:} String Gas
Cosmology; Dilaton - Gravity; Closed String; Thermodynamics.
\end{abstract}
\section{Introduction}
String gas cosmology [1] is one of the best model of the early
universe, which is called Brandenberger - Vafa (BV) scenario. This
model claimed that the early universe was small and compact and
surrounded by hot and high dense string gas. Because of vanishing
winding modes of strings, (A closed string can wrap around the extra
dimension, such a state is called a winding mode and is appropriate
with inverse of extra dimension, ie. $\omega\sim1/R$), three
dimensions of space become large ($\omega\rightarrow0 \Rightarrow
R\rightarrow\infty$). For more studying one may see several papers
[2-13]. One of the important problem in this model is consideration
of thermal equilibrium in the early universe. In that case the early
time cosmic evolution in string gas cosmology considered by means of
open strings attached to $D$-branes [14]. In that paper, statistical
properties of open strings in $D$-brane background reviewed and
string fields determined by using dilaton - gravity equations in the
Hagedorn regime. Authors in Ref. [14] concluded that initial
conditions can be fine tuned to have thermal equilibrium in the
beginning, but, in a short time, string gas falls out of thermal
equilibrium. Through the similar method of Ref. [14] dominated by
closed strings, we study thermal equilibrium in string gas
cosmology. In this way we perform
different calculation with the Ref. [15] about the interaction rate.\\
In the Ref. [16] some aspects of string gas cosmology at finite
temperature described for two cases, (I) Hagedorn regime in a very
small homogeneous and isotropic universe, (II) a radiation regime
with two independent scale factors corresponding to large and small
dimensions. In the radiation regime the lightest Kaluza - Klein and
winding mode contributions considered. It is known that in both
Hagedorn and radiation regimes some matters have manifestly
$T$-duality invariant. Importance of the radiation regime explained
well in the Ref. [16], therefore we are interest to consider this
regime to
study of thermal equilibrium in the early universe.\\
As we told, Open strings and Hagedorn regime in string gas cosmology
studied [14]. Here, we follow same way and consider thermodynamics
of string gas in the early universe which is dominated by the closed
strings. We study thermal equilibrium condition for the early time
of the universe in Hagedorn and radiation regime. We take equations
of motion from Ref. [16] and solve them in the way of Ref. [14] and
explain requirements of thermal equilibrium.\\
This paper organized as the following. In the section 2 we review
thermodynamics of closed strings and write the relevant entropy for
the small and large radius. Then in the section 3 we consider
dilaton - gravity in the Hagedorn regime with a single scale factor,
and study thermal equilibrium by calculating Hubble parameter and
the interaction rate. In the section 4 we repeat analysis of the
section 3 in the almost-radiation regime and with two scale factors.
Finally in the section 5 we give conclusion and discussion of the
thermal equilibrium in the early universe.
\section{The closed string entropy}
In this section we review the calculation of the closed strings
entropy in dilaton - gravity background. We assume that all of
directions are small and compact, which represented by a flat
$T^{9}$ torus. Also we would like to consider our system at
temperature close to the Hagedorn temperature and in low - energy
limit. Therefore there are $D=9$ compact coordinates which are
represented by $x_{i}$ with $i=1,2...9$. The space - time metric and
dilaton field may be written as the following,
\begin{eqnarray}\label{s1}
ds^{2}&=&-dt^{2}+R_{i}^{2}dx^2,\nonumber\\
R_{i}&=&e^{\lambda_{i}(t)},\nonumber\\
\phi&=&\phi(t),
\end{eqnarray}
where $R_{i}$ $(i=1,2...,9)$ denote the scale-factor of the torus
and are equal to the corresponding radii for the world - volume. So,
one can define the volume of the space by $V=(R_{i})^D$. Because of
the string T-duality ($R\rightarrow 1/R$ symmetry ) we can take all
dimensions to be larger than string scale, ie. $R_{i}\geq1$,
moreover we set $\alpha^{\prime}=1$ for convenient. By consideration
of the closed string model we want to study the early time cosmic
evolution in string gas cosmology. In the microcanonical ensemble
there is a critical temperature which is called the Hagedorn
temperature [17], where the partition function of a free string gas
diverges. Therefore in this regime the usual thermodynamical
equivalence between the canonical and microcanonical ensembles can
break down and more fundamental ensemble must be used. We use
different methods from Ref. [16] to obtain the entropy. By using
density of single string state [18] we will obtain thermodynamics of
the closed strings. The closed string energy $\varepsilon$
corresponds to the length of random walk and the number of them grow
with factor $\exp(\beta_h\varepsilon)$, where $\beta_{h}$ is inverse
of the Hagedorn temperature. Also there is a volume factor of random
walk $V_{walk} = W$. In addition there is a factor $V$ to
transformation of zero state and a factor $1/\varepsilon$ because of
periodic nature of the closed strings. In order to specify the
single string density of states we must account all of above factors
which yield to the following expression,
\begin{equation}\label{s2}
\omega(\varepsilon)\sim\frac{V\exp(\beta_{h}\varepsilon)}{\varepsilon
W}.
\end{equation}
By using the T-duality in string theory, the small and compact space
has dual picture as non-compact space, hence we
consider two following cases:\\
(i) $R_{i}\ll\sqrt{\varepsilon}$, it is small radius regime and all
of directions are effectively compact which is called space-filling.
In this case one can find
$\omega(\varepsilon)=\frac{\exp(\beta_{h}\varepsilon)}{\varepsilon}$.\\
(ii) $R_{i}\gg\sqrt{\varepsilon}$, it is large radius regime and all
of directions are effectively non-compact which is called
well-contained. In this case one can find,
$\omega(\varepsilon)={V\exp(\beta_{h}\varepsilon)}{\varepsilon^{D^{\prime}}}$,
where $D^{\prime}=-\frac{D+1}{2}$. In the small radius regime, where
$\varepsilon\gg1$, the value of energy is sufficient to excite
winding modes of closed string, but in the large radius regime,
where $\varepsilon\ll1$, the value of energy is lower than
excitation of winding modes. Then the total density of states,
$\Omega(E)$, is given by [14],
\begin{equation}\label{s3}
\Omega(E)=
\Sigma_{k}\frac{1}{k!}\prod_{i=1}^{k}\int_{0}^{E}\Omega(\varepsilon_{i})d\varepsilon_{i}\delta(\Sigma{\varepsilon_{i}}-E),
\end{equation}
where $k$ is the number of strings. The equation (3) can be
rewritten in the form of the following expression,
\begin{eqnarray}\label{s4}
\Omega(E)=\frac{1}{2\pi E}\int_{-\infty}^{\infty}d\alpha
e^{-i\alpha}e^{F(\alpha)},
\end{eqnarray}
where $F(\alpha)$ is defined as,
\begin{eqnarray}\label{s5}
F(\alpha)=\int_{0}^{E}
d\varepsilon\omega(\varepsilon)e^{\frac{i\alpha\varepsilon}{E}}.
\end{eqnarray}
We must note that $F(\alpha)$ is a regular function which vanishes
at $\alpha\rightarrow\pm\infty$ limit, and therefore the integral
(4) is not divergent. In order to obtain entropy we calculate the
integral (4) in both small and large radius. In the first case,
small radius, one can obtain $F(\alpha)=\ln E+\gamma E$, where
$\gamma=\frac{\beta_{h}E+i\alpha}{E}$. Therefore entropy is is given
by $S=ln \Omega(E)\simeq\beta_{h}E$. Similarly in the second case
for large radius, one can obtain,
\begin{equation}\label{s6}
S=\beta_{h}E+\ln\frac{z^n}{n!(n+D^{\prime})!}
\end{equation}
where, $z=-(-1)^{D^{\prime}}D^{\prime}!VE^{D^{\prime}}$, and for
$D=9$ it reduces to $z=-120VE^{-5}$. In the next sections we use
above information to study the thermal equilibrium of a given
system.
\section{Dilaton gravity and Hagedorn regime}
In this section we shall study the dilaton - gravity equations of
motion with a massless dilaton field $\Phi$ corresponding to the
low-energy effective action of string theory in $D+1$ space-time
dimension [14,19] at weak string coupling, which is described by the
following action,
\begin{equation}\label{s7}
S=\int d^{D+1}
x\sqrt{-g}\left[e^{-2\Phi}(R+4(\nabla\Phi)^2)+{\mathcal{L}}_{M}\right],
\end{equation}
where $g$ is the determinant of the background metric $g_{\mu\nu}$,
and ${\mathcal{L}}_{M}$ denotes the lagrangian of some matter. The
coupling of dilaton field with gravity is convenient in string
theory. Here, if we consider spatially homogeneous field
configuration, then the action (7) exhibits a low energy
manifestation of the string T-duality symmetry [14]. By introducing
a shifted dilaton $\psi=2\Phi-\lambda_{i}$, $(i=1,2,...,9)$, one can
simplify the equations of motions of the dilaton - gravity system as
the following,
\begin{eqnarray}\label{s8}
-d\dot{\mu}^{2}-(9-d)\dot{\nu}^{2}+\dot{\psi}^2&=&Ee^{\psi},\nonumber\\
\ddot{\mu}-\dot{\mu}\dot{\psi}&=&\frac{1}{2}P_{d}e^{\psi},\nonumber\\
\ddot{\nu}-\dot{\nu}\dot{\psi}&=&\frac{1}{2}P_{9-d}e^{\psi}\nonumber\\
\ddot{\psi}-d{\dot{\mu}^2}-(9-d)\dot{\nu}^2&=&\frac{1}{2}Ee^{\psi},
\end{eqnarray}
where derivatives are with respect to the cosmic time $t$. Also
$P_{d}=-\frac{\partial F}{\partial \mu_i}$ with $i=1,2,...,d$, and
$P_{9-d}=-\frac{\partial F}{\partial \nu_i}$ with $i=d+1,...,9$, are
given in terms of the free energy $F$. In the equation (8), $E$
denotes the total energy found by multiplying the total spatial
volume of the space by the energy density appearing in
$\mathcal{L}_m$ of action (7). Also in the equation (8) we assumed
that the background is homogeneous and isotropic in $d$-spatial
dimensions and $(9-d)$-small spatial dimensions. We denoted the
large and small dimensions with their corresponding scale factors as
$R=\exp(\mu)$ and $r=\exp(\nu)$. In contrast with evolution
equations of open strings there is not energy density $\rho$ in the
right hand side of the second and third equations (8). It means that
the force along the corresponding directions, which determines the
cosmic evolution, is only given by the pressures, and not sum of the
pressure with energy density. By using entropy, which is obtained in
the previous section, one can find the temperature and the pressures
as,
\begin{eqnarray}\label{s9}
\frac{1}{T}&=&\frac{\partial S}{\partial E},\nonumber\\
P_{d}&=&T V_{d}\frac{\partial S}{\partial V_{d}},\nonumber\\
P_{9-d}&=&T V_{9-d} \frac{\partial S}{\partial V_{9-d}},
\end{eqnarray}
so, $V=V_{d}V_{9-d}$. Therefore one can find
$\frac{1}{T}=\beta_{h}$, $P_{d}=P_{9-d}=0$ and
$\frac{1}{T}=\beta_{h}+\frac{D^{\prime}}{E}$,
$P_{d}=P_{9-d}=\frac{nE}{\beta_{h}E+D^{\prime}}$ for small and large
radius, respectively, where $n$ is a large integer number. It shows
that for small radius regime, temperature is equal to the Hagedorn
temperature and pressures are zero, but for large radius regime,
temperature is always smaller than the Hagedorn temperature and
pressures are very large.\\
Now, in two following subsection, we are going to discuss the small
and large radius.
\subsection{small radius}
As we see, in the small radius regime, pressures are zero, thus one
can set $\mu=\nu$, so the equations (8) reduce to,
\begin{eqnarray}\label{s10}
{\dot{\psi}^2}-9\dot{\mu}&=&Ee^{\psi},\nonumber\\
\ddot{\mu}-\dot{\mu}\dot{\psi}&=&0,\nonumber\\
\ddot{\psi}-9{\dot{\mu}^2}&=&\frac{1}{2}Ee^{\psi}.
\end{eqnarray}
The second equation of (10) gives the following condition,
\begin{equation}\label{s11}
\dot{\mu}=Ce^{\psi},
\end{equation}
where $C$ is the integration constant. We assume that initial values
of fields are as $\psi(0)=\psi_{0}$, $\dot{\psi}(0)=\dot{\psi}_{0}$,
$\mu(0)=\mu_{0}$ and $\dot{\mu}(0)=\dot{\mu}_{0}$, therefore the
constant of integral in the relation (11) fixed as
$C=\dot{\mu}_{0}e^{-\psi_{0}}$. From the condition (11) we see that,
when $\dot{\mu}_{0}$ is positive the expansion rate for the small
dimensions is always positive and vis versa. Also from the first and
third equation of (10) and condition (11) we can find,
\begin{eqnarray}\label{s12}
\psi&=&-\ln{[\frac{E}{4}t^{2}-C_{1}t+C_{2}]},\nonumber\\
\mu&=&\frac{1}{3}\ln{\left[\frac{Et-2(C_{1}+3C)}{Et-2(C_{1}-3C)}\right]}+\mu_{0},
\end{eqnarray}
where $C_{2}=e^{-\psi_{0}}$ and $C_{1}=\sqrt{E C_{2}+9C^{2}}$.\\
Now, Hubble expansion parameter can be determined as,
\begin{equation}\label{s13}
H=\dot{\mu}=\frac{C}{\frac{E}{4}t^{2}-C_{1}t+C_{2}},
\end{equation}
so, the initial Hubble rate is $H_{0}=\dot{\mu}_{0}$. We draw the
plots of functions $R=e^{\mu}$, $H=\dot{\mu}$ and $\psi$ for
$C_{1}=395.401931$, $C_{2}=148.413159$ and $C=29.6826318$ in figure
1 (We take initial values of fields from the Ref. [16]).

\begin{figure}[th]
\begin{center}
\includegraphics[scale=.29]{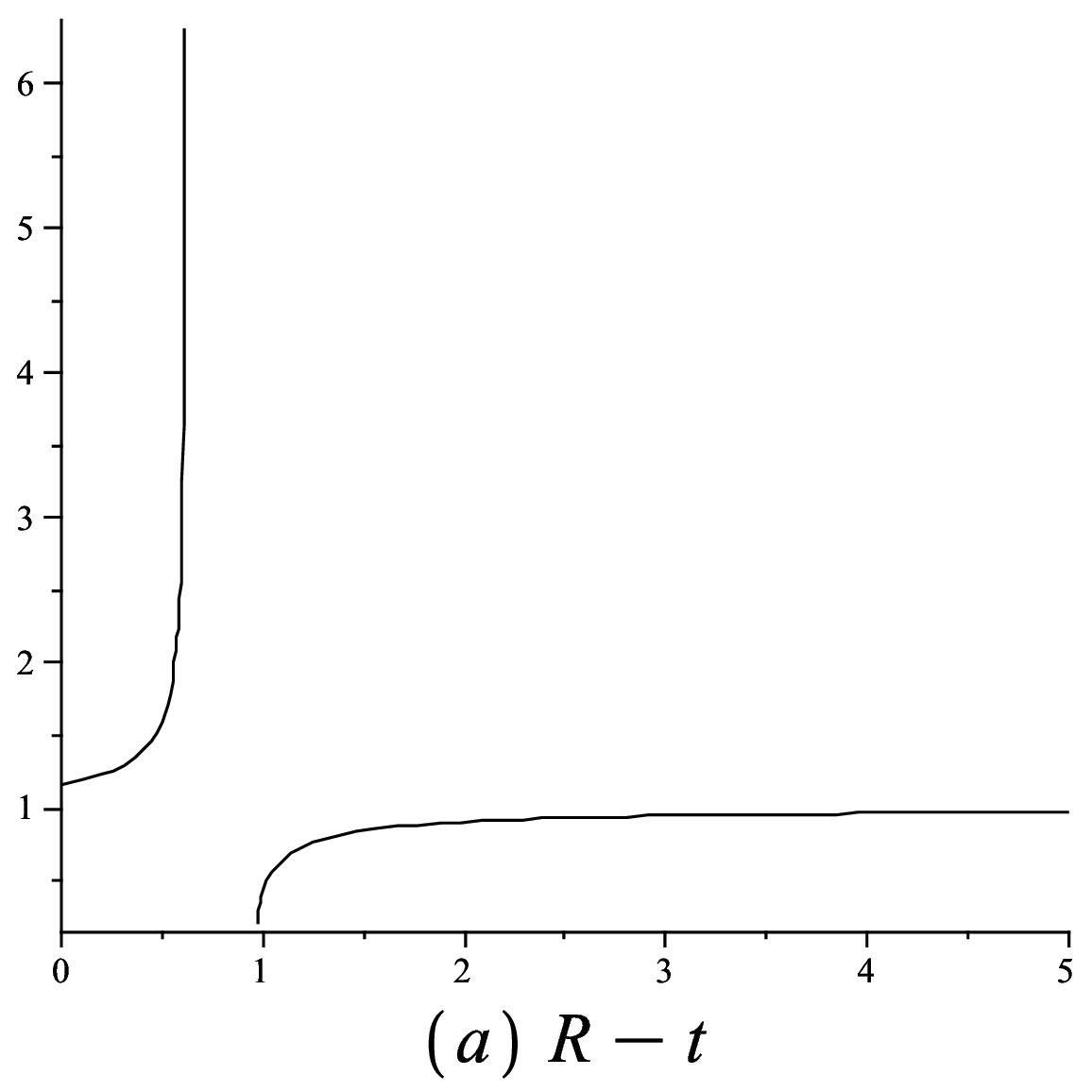}\includegraphics[scale=.29]{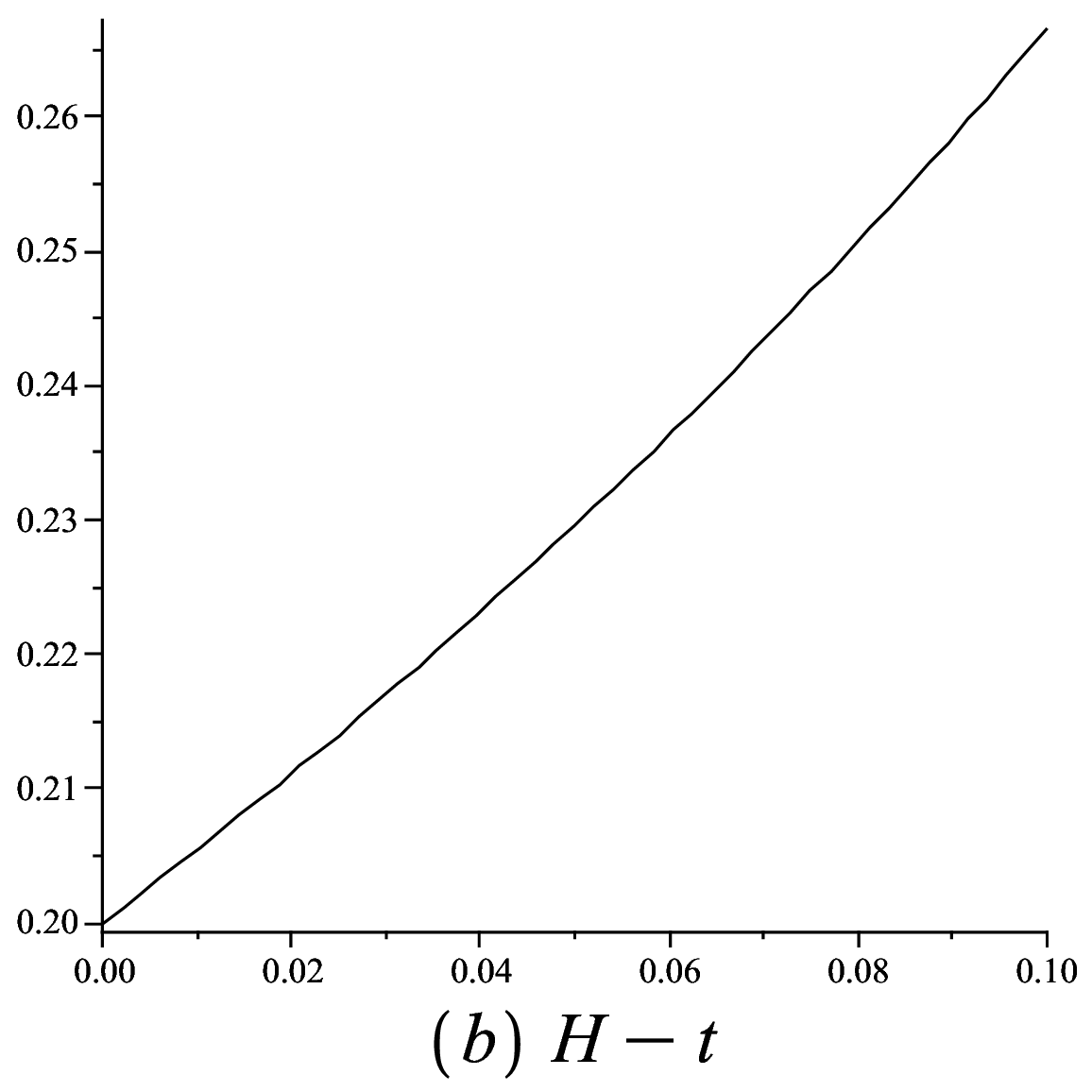}\includegraphics[scale=.25]{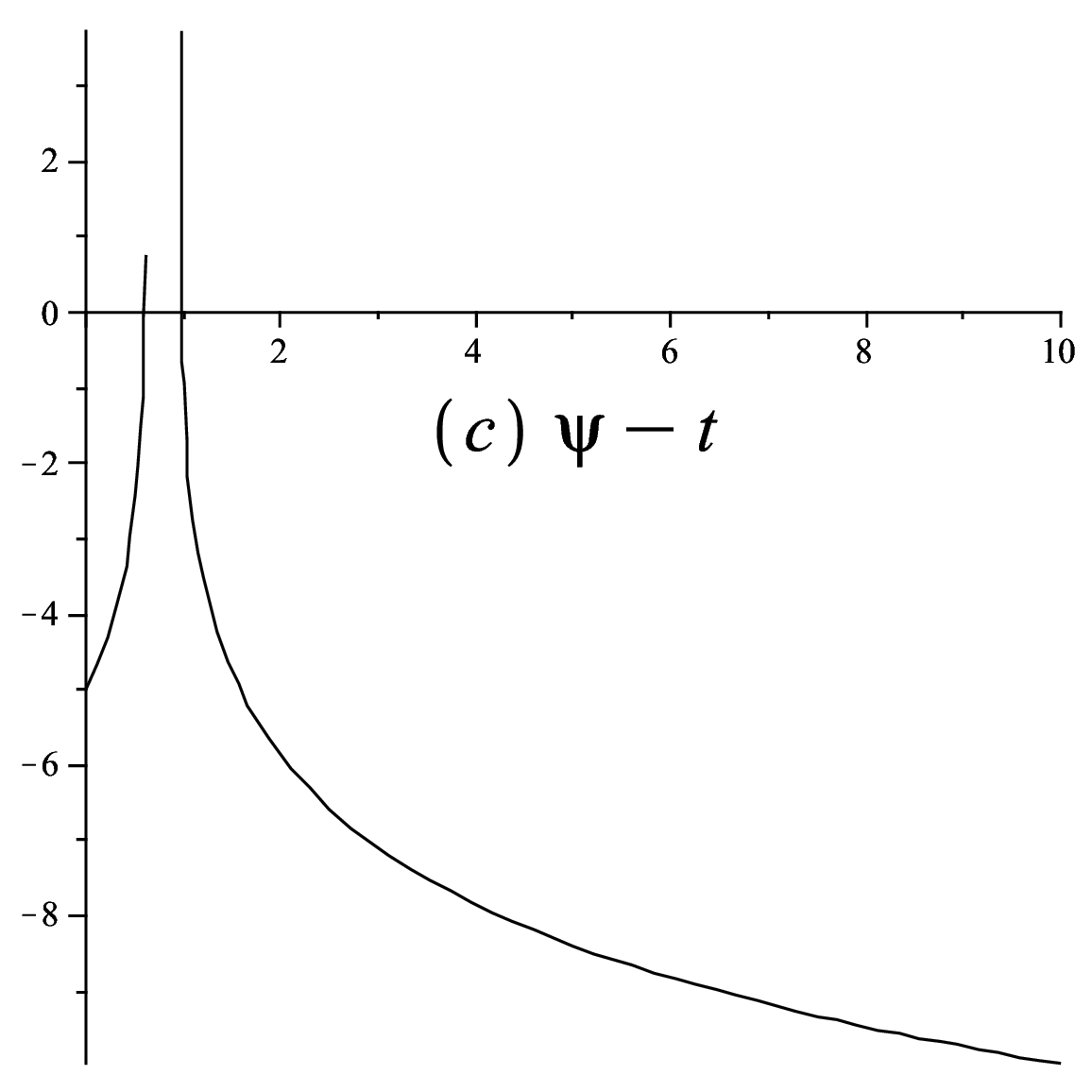}
\caption{(a) plot of scale factor $R$ in terms of cosmic time $t$
for the small radius of Hagedorn regime. (b) plot of Hubble
parameter $H$ in terms of cosmic time $t$ for the small radius of
Hagedorn regime. (c) plot of dilaton field $\psi$ in terms of cosmic
time $t$ for the small radius of Hagedorn regime. We take initial
values $\mu_{0}=0.001$, $\dot{\mu}_{0}=0.2$ and $\psi_{0}=-5$ from
the Ref. [16].}
\end{center}
\end{figure}

Then we must evaluate the interaction rate per string, $\Gamma$. We will calculate interaction rate for two cases, short and large strings. Before that we
should note that in the Ref. [15] string interaction rates in the string gas cosmology has been studied. We use the idea of Ref. [14] to divide interaction
rates to short and long string cases, then we use corresponding values of $\omega(\varepsilon)$ and $\Omega(E)$ for the closed string case, which
introduced in the section 2. In that case one can find $\Gamma_{short}\sim e^{\psi}[-1+\ln{E}]$ and $\Gamma_{long}\sim
\frac{-1+\ln{E}}{2\ln{E}}E^{2}e^{\psi}$. Now thermal equilibrium requires that the interaction rate per string $\Gamma$ to be larger than the expansion
rate $H$: $\Gamma\geq H$. This condition satisfied for short strings if $\ln{E}\geq C$. In another word if $\ln{E}\geq \dot{\mu}_{0}e^{-\psi_{0}}+1$ there
is thermal equilibrium for short strings. For the initial values of field which given by Ref. [16] there is not thermal equilibrium. In another word, with
$E\geq e^{29}$ there is thermal equilibrium for short strings.
It means that, very large amount of energy requested  to have thermal equilibrium in presence of short strings.\\
Also there is thermal equilibrium for long string if
$\frac{E^{2}}{2}(1-(\ln{E})^{-1})\geq C$. The initial values of
fields say that there are thermal equilibrium for long strings in
small radius of Hagedorn regime.\\
There is important discussion about above results. It is known that
thermodynamics should be used to determine whether the system is
dominated by short or long strings. We just written possible
conditions to have thermal equilibrium in presence of the short and
long strings.
\subsection{large radius}
In the section 2 we have shown that for large radius
$P_{d}=P_{d-9}$. Now we denote pressures with $P$ and rewrite
equations of motion (8) as following,
\begin{eqnarray}\label{s14}
{\dot{\psi}^2}-9\dot{\mu}^{2}&=&Ee^{\psi},\nonumber\\
\ddot{\mu}-\dot{\mu}\dot{\psi}&=&\frac{1}{2}Pe^{\psi},\nonumber\\
\ddot{\psi}-9{\dot{\mu}^2}&=&\frac{1}{2}Ee^{\psi}.
\end{eqnarray}
It is clear that when the pressures are not ignorable the energy-momentum conservation requires both $E$ and $P$ to depend on time. In the case of
zero-pressure energy-momentum conservation imposes the energy $E$ to be a constant. The Ref. [20] pointed out these discusses solutions with non-vanishing
pressure, and showed that it is very difficult to obtain explicit solutions when pressures are very large. Now, we suppose that pressure $P$ is
approximately constant and one can obtain solutions of equation (14) as following expressions,
\begin{eqnarray}\label{s15}
\psi&=&-\ln{[\frac{E}{4}t^{2}-C_{1}t+C_{2}]},\nonumber\\
\mu&=&\frac{1}{3}\ln{\left[\frac{E(\frac{E}{2}+\frac{3P}{8})t^{2}-C_{1}(2E+\frac{3P}{2})t+2C_{1}^{2}
+\frac{3C_{2}P}{2}}{E(\frac{E}{2}-\frac{3P}{8})t^{2}-C_{1}(2E-\frac{3P}{2})t+2C_{1}^{2}-\frac{3C_{2}P}{2}}\right]},
\end{eqnarray}
where $C_{2}=e^{-\psi_{0}}$ and
$C_{1}=\sqrt{\frac{3PC_{2}}{4}\frac{e^{3\mu_{0}}+1}{e^{3\mu_{0}}-1}}$.
Therefore one can find Hubble parameter as,
\begin{equation}\label{s16}
H=-\frac{P}{2Et-4C_{1}},
\end{equation}
so, the initial Hubble rate is $H_{0}=\frac{P}{4C_{1}}$. We draw the plots of functions $R=e^{\mu}$, $H=\dot{\mu}$ and $\psi$ for $C_{1}=27238.6522$,
$C_{2}=148.413159$ and $P=10^{4}$ in figure 2 (We take initial values of fields from Ref. [16]). As we can see from Fig.s 2 (a) and  (b) there is Jeans
instability for the case of closed strings which is agree with Ref. [14].

\begin{figure}[th]
\begin{center}
\includegraphics[scale=.29]{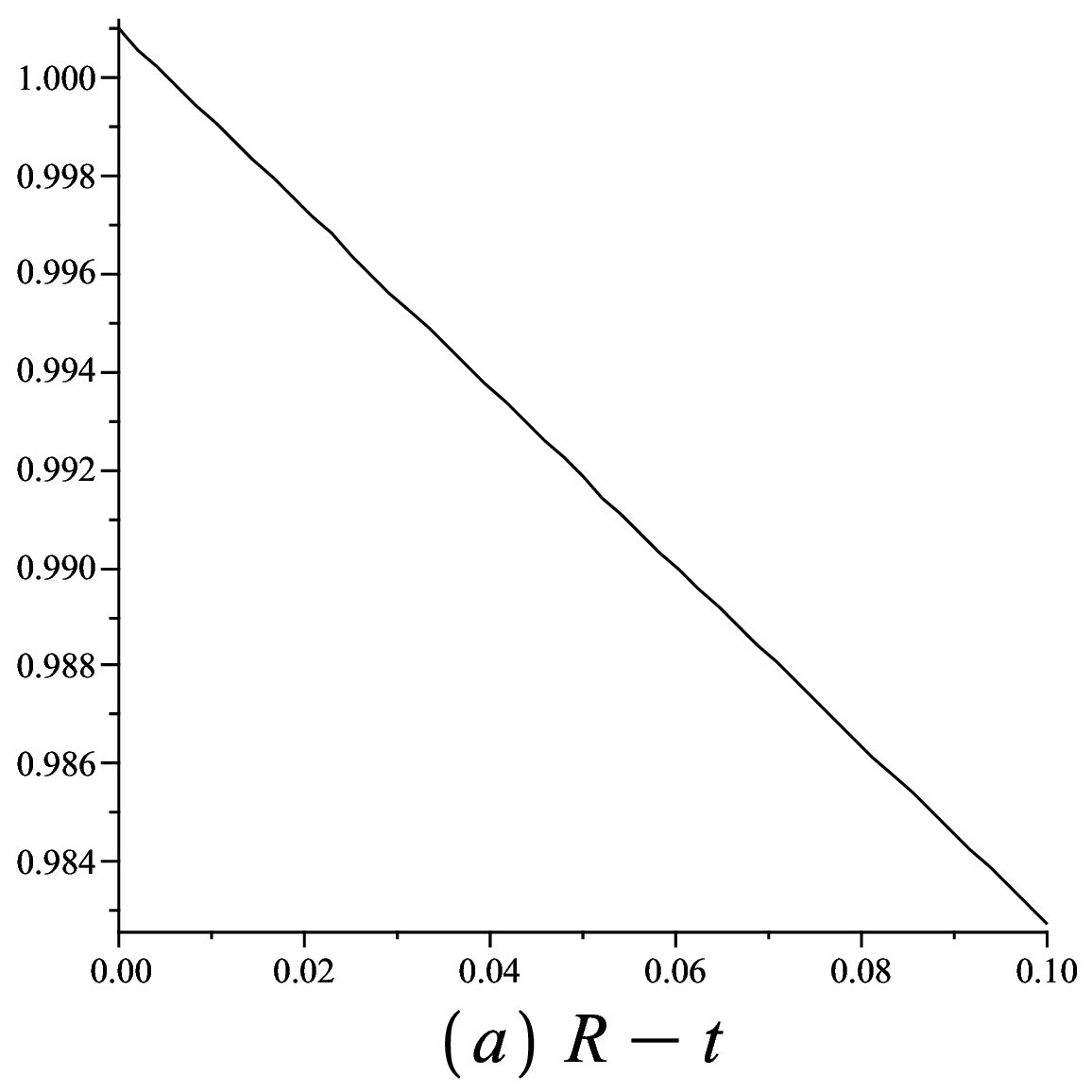}\includegraphics[scale=.29]{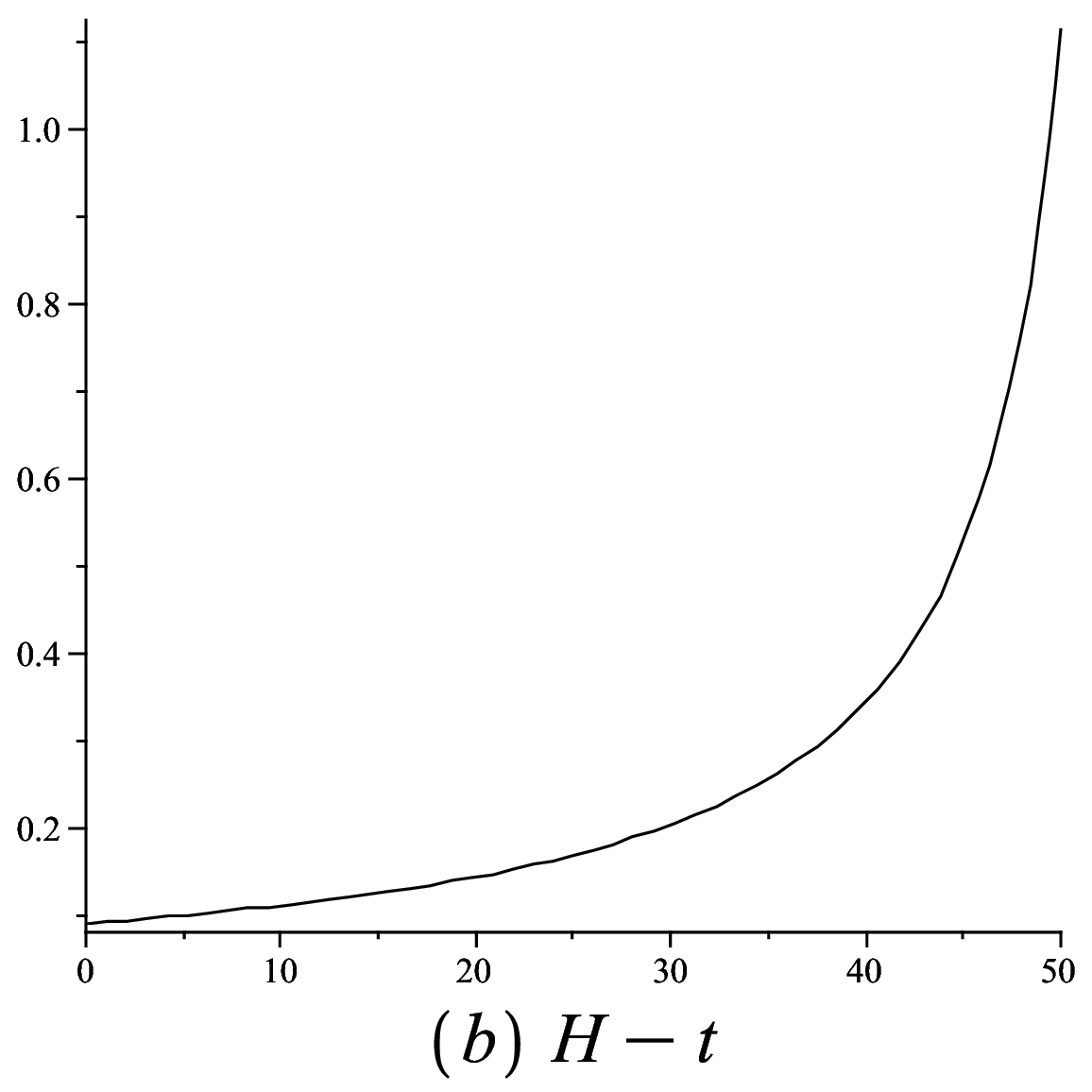}\includegraphics[scale=.25]{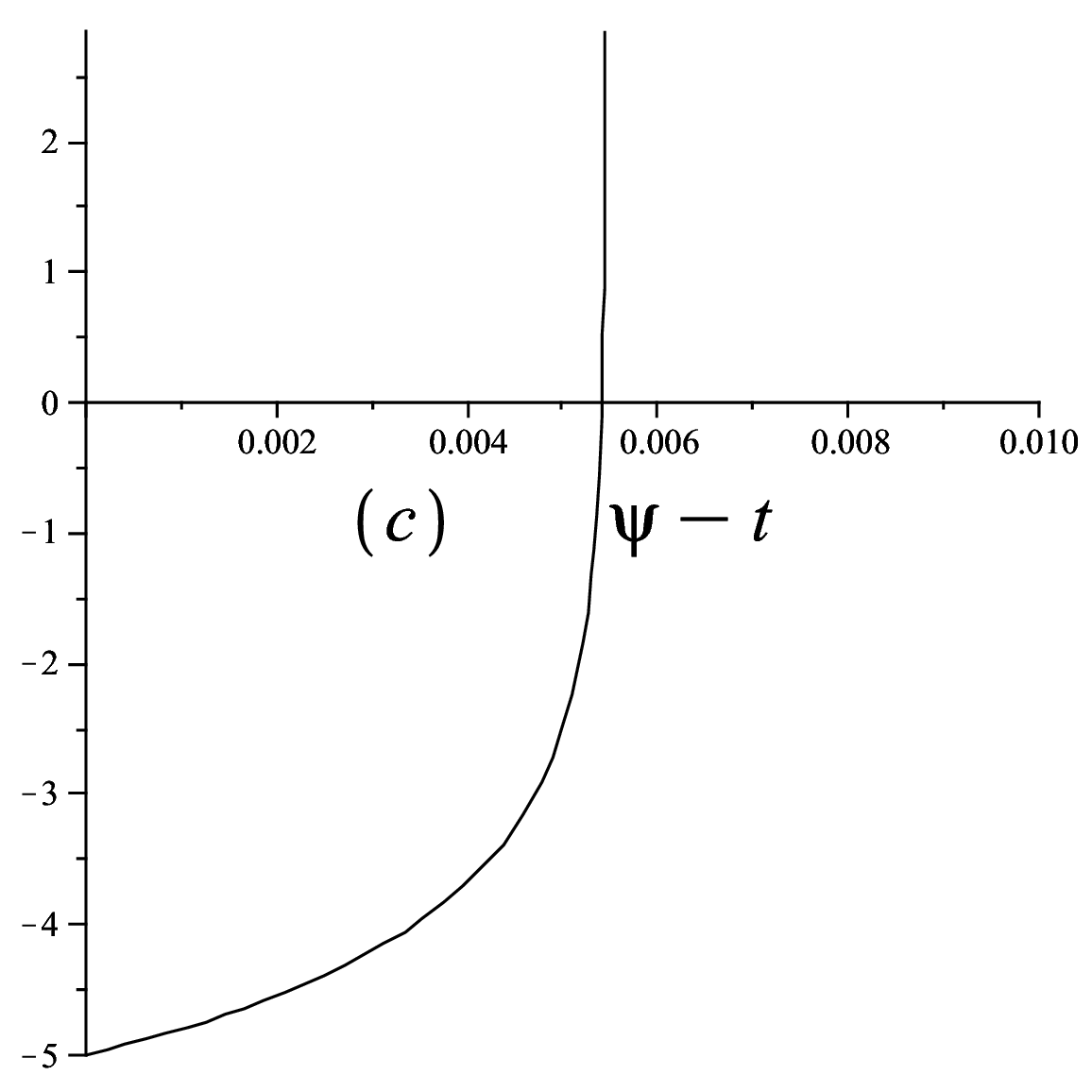}
\caption{(a) plot of scale factor $R$ in terms of cosmic time $t$ for the large radius of Hagedorn regime. (b) plot of Hubble parameter $H$ in terms of
cosmic time $t$ for the large radius of Hagedorn regime. (c) plot of dilaton field $\psi$ in terms of cosmic time $t$ for the large radius of Hagedorn
regime. We take initial values $\mu_{0}=0.001$, $\dot{\mu}_{0}=0.2$ and $\psi_{0}=-5$ from the Ref. [16].}
\end{center}
\end{figure}

In order to obtain thermal equilibrium conditions (which is a relation between $P$ and $E$), we must have $8(2C_{1}-Et)\ln{E}\geq(Et^{2}-4C_{1}+4C_{2})P$
for short strings, and also have $4E^{2}(1-\frac{1}{\ln{E}})(2C_{1}-Et)\geq(Et^{2}-4C_{1}+4C_{2})P$
for long strings.\\
Now, let us rewrite equations (14) as the following form,
\begin{eqnarray}\label{s17}
{\psi}^{\prime\prime}&=&-\frac{1}{2}Ee^{-\psi},\nonumber\\
\mu^{\prime\prime}&=&\frac{1}{2}Pe^{-\psi},\nonumber\\
{\psi^{\prime}}^{2}-9{\mu^{\prime}}^2&=&Ee^{-\psi},
\end{eqnarray}
where we used conformal time $\eta$ given by $e^{-\psi}d\eta=dt$. In
the equations (14) derivatives are with respect to $\eta$. The first
two equations of (14) give the following relation between $\psi$ and
$\mu$,,
\begin{equation}\label{s18}
\psi=-\frac{E}{P}\mu+(\frac{E}{P}\mu_{0}^{\prime}+\psi_{0}^{\prime})\eta+\frac{E}{P}\mu_{0}+\psi_{0}.
\end{equation}
Under assumption of $E\ll P$ and using relations (17) and (18), one
can find following expressions for $\psi$ and $\mu$ in terms of
conformal time $\eta$,
\begin{eqnarray}\label{s19}
\psi&=&-\frac{E}{2{\psi_{0}^{\prime}}^{2}}e^{-\psi_{0}}(e^{-\psi_{0}^{\prime}\eta}-1)
+(\psi_{0}^{\prime}-\frac{E}{2\psi_{0}^{\prime}}e^{-\psi_{0}})\eta+\psi_{0},\nonumber\\
\mu&=&+\frac{P}{2{\psi_{0}^{\prime}}^{2}}e^{-\psi_{0}}(e^{-\psi_{0}^{\prime}\eta}-1)
+(\mu_{0}^{\prime}+\frac{P}{2\psi_{0}^{\prime}}e^{-\psi_{0}})\eta+\mu_{0}.
\end{eqnarray}
So, we find Hubble parameter as $H=\dot{\mu}=\mu^{\prime}\exp{\psi}$. In figure 3 we give plots of $R$, $H$ and $\psi$ in terms of conformal time $\eta$,
with initial values of $\psi_{0}=-5$, $\psi_{0}^{\prime}=395.403377$, $\mu_{0}=0.001$, $\mu_{0}^{\prime}=29.6826318$, $E=1000$ and $P=10^{4}$ [16]. In
contrast with previous case, which is expressed in terms of cosmic time, we can see from Fig.s 3 (a) and (b) that, there is Jeans stability in conformal
time for closed strings.

\begin{figure}[th]
\begin{center}
\includegraphics[scale=.29]{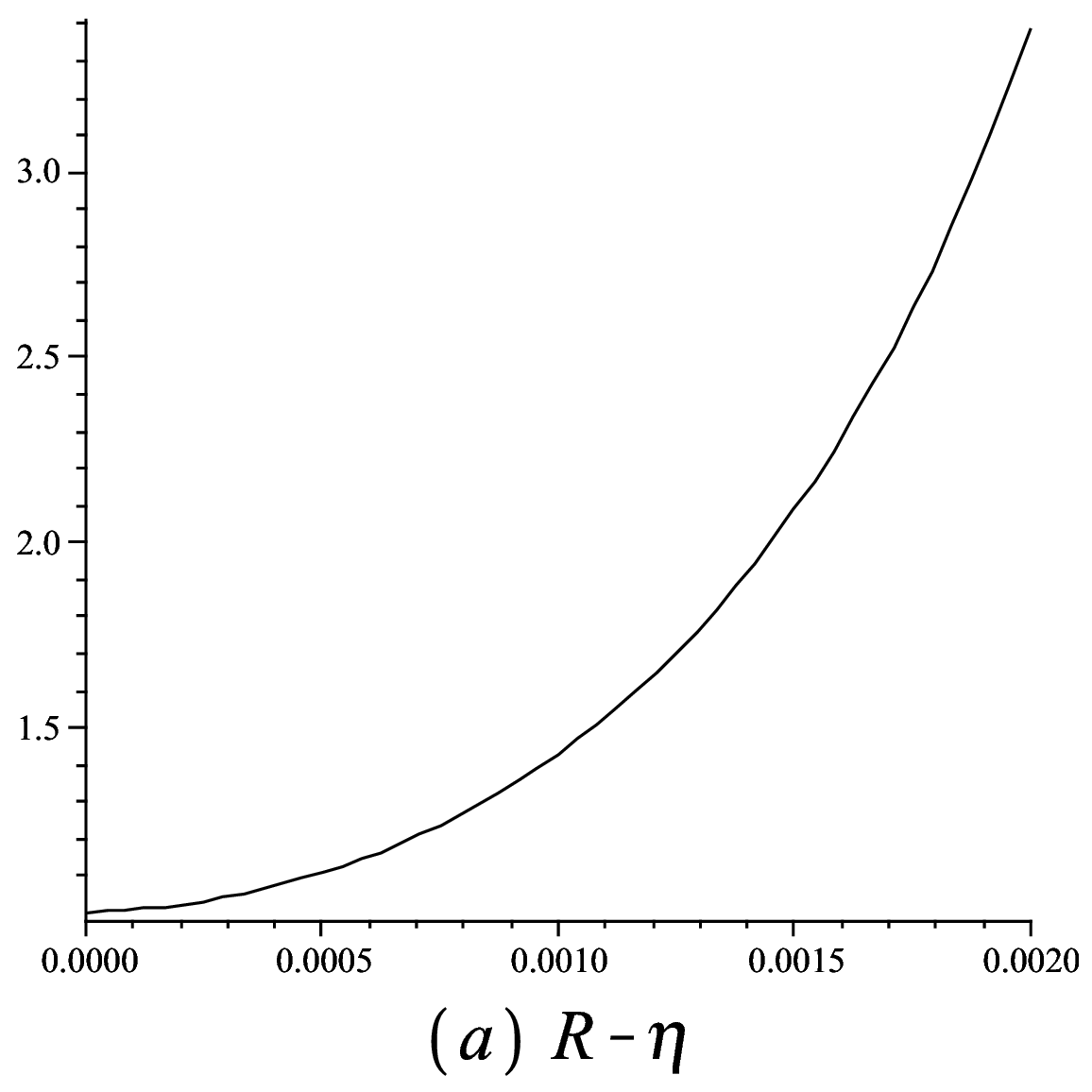}\includegraphics[scale=.29]{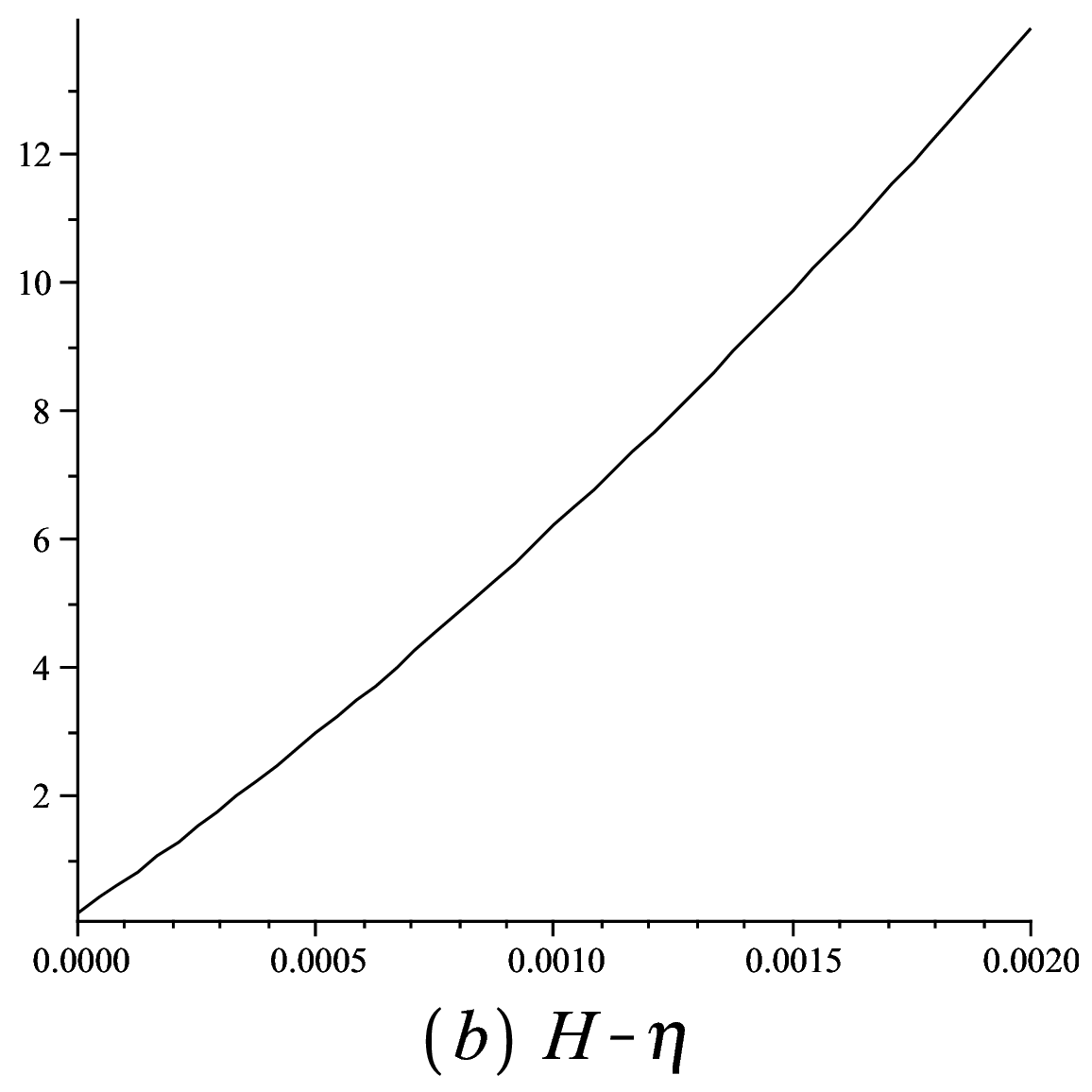}\includegraphics[scale=.25]{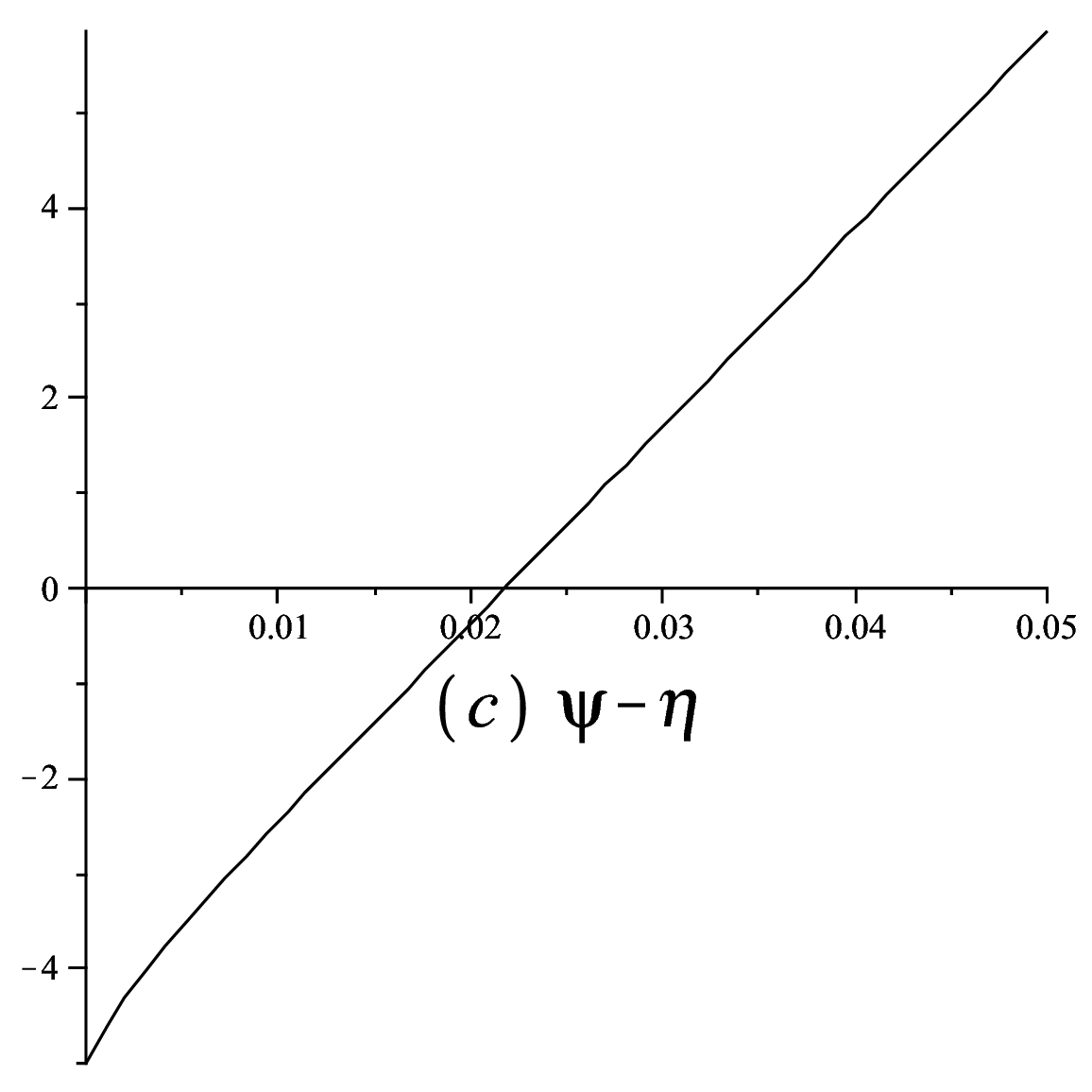}
\caption{(a) plot of scale factor $R$ in terms of conformal time $\eta$ for the large radius of Hagedorn regime. (b) plot of Hubble parameter $H$ in terms
of conformal time $\eta$ for the large radius of Hagedorn regime. (c) plot of dilaton field $\psi$ in terms of conformal time $\eta$ for the large radius
of Hagedorn regime.We take initial values $\mu_{0}=0.001$, $\dot{\mu}_{0}=0.2$ and $\psi_{0}=-5$ from the Ref. [16].}
\end{center}
\end{figure}

Comparing the Hubble parameter with the interaction rate for short and long string yields us to following thermal equilibrium conditions for short and long
strings respectively,
\begin{equation}\label{s20}
\ln
E\geq-\frac{P}{2\psi_{0}^{\prime}}e^{-\psi_{0}}(e^{-\psi_{0}^{\prime}\eta}-1)+\mu_{0}^{\prime},
\end{equation}
and
\begin{equation}\label{s21}
\frac{E^{2}}{2}(1-\frac{1}{\ln{E}})\geq
-\frac{P}{2\psi_{0}^{\prime}}e^{-\psi_{0}}(e^{-\psi_{0}^{\prime}\eta}-1)+\mu_{0}^{\prime}.
\end{equation}
In figure 4 we describe thermal equilibrium conditions for short and long strings. In the Fig. 4 (a) curve of $\ln{E}$ drawn for $E\sim 10^{13}$ and in
Fig. 4 (b) curve of $\frac{E^{2}}{2}(1-\frac{1}{\ln{E}})$ drawn for small energy ($E\in[4,20]$). On the other hand in Fig. 4 (c) curve of right hand sides
of equations (21) and (22) represented which have linear behavior for early time. From Fig.s 4 (a) and 4 (c) one can find that there is thermal equilibrium
in the early universe if and only if the value of energy became very large ($E\sim 5\times10^{13}$). Therefore under consideration of initial values of
Ref. [16] ($E\in[500, 5000]$) there is not thermal equilibrium for short strings. On the other hand, by comparing Fig.s 4 (b) and 4 (c), one can see that
thermal equilibrium condition in presence of long strings satisfied at the early universe with low energy ($E\sim10$). Again, for initial value where
$E\in[500, 5000]$ there is not thermal equilibrium for short strings in Hagedorn regime.

\begin{figure}[th]
\begin{center}
\includegraphics[scale=.29]{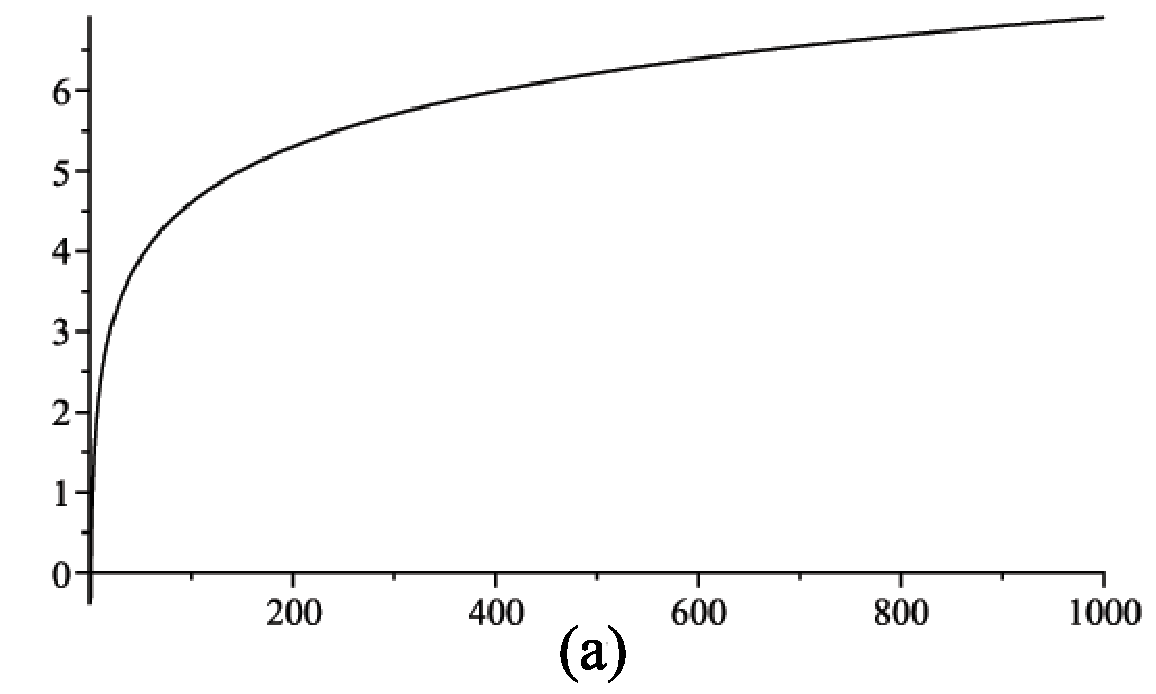}\includegraphics[scale=.29]{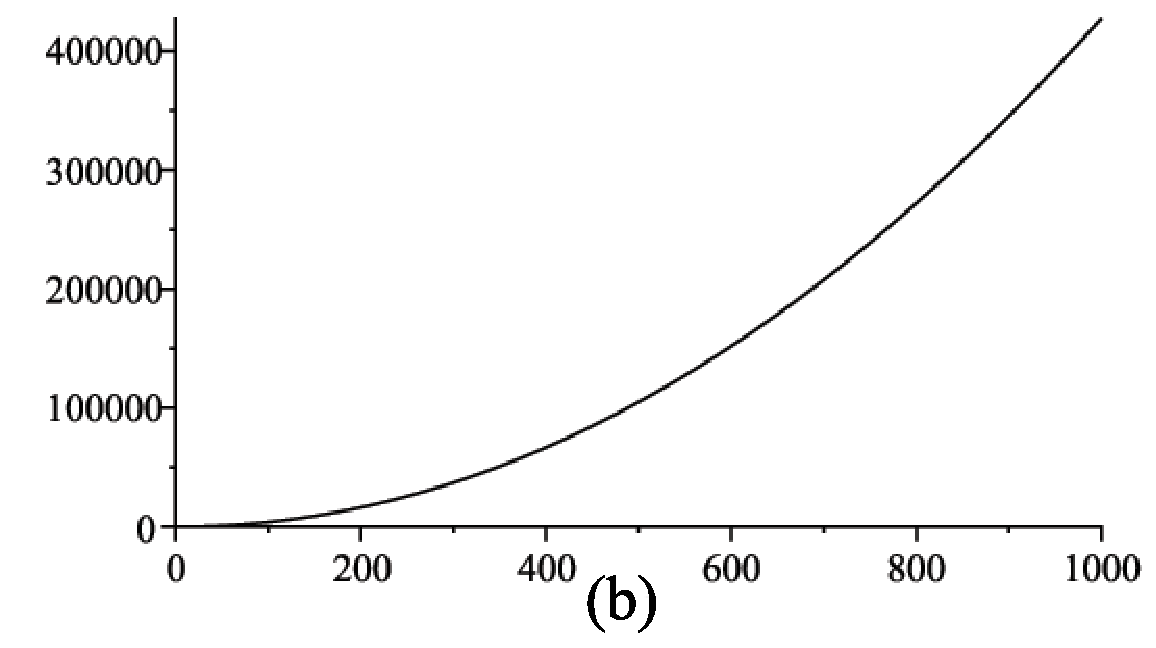}\includegraphics[scale=.25]{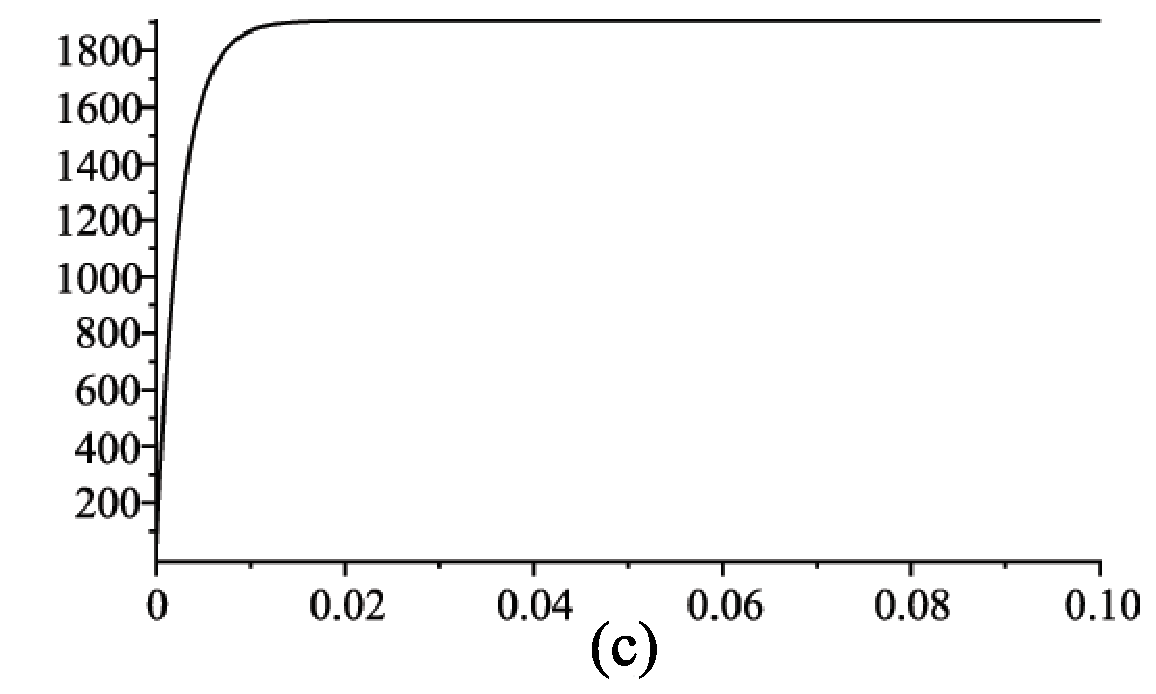}
\caption{(a) plot of $\ln E$ for short strings in Hagedorn regime. (b) plot of $\frac{E^{2}}{2}(1-\frac{1}{\ln{E}})$ for long strings in Hagedorn
regime.(c) plot of $\mu_{0}^{\prime}-\frac{P}{2\psi_{0}^{\prime}}e^{-\psi_{0}}(e^{-\psi_{0}^{\prime}\eta}-1)$ in Hagedorn regime. We take initial values
$\psi_{0}=-5$, $\psi_{0}^{\prime}=395.403377$ and $\mu_{0}^{\prime}=29.6826318$ [16]. }
\end{center}
\end{figure}

\section{Dilaton gravity and pure radiation regime}
In this section, again we consider action (7) and equations of
motion (8). We assume that $d$ dimensions ($R$) start to expand
while $9-d$ dimensions ($r$) remain small [16]. In this procedure
the temperature reach below the Hagedorn regime where dynamics of
the system described by massless states which is called radiation
regime. We would like to solve the dilaton - gravity equations (8)
and dedermine all fields. Then by specifying Hubble parameter and
interaction rate we are able to study thermal equilibrium of the
system. As we can see from Ref. [16],
$P_{rad}^{(d)}=\frac{E_{rad}^{(d)}}{d}$ and $P_{rad}^{(9-d)}=0$. So,
one can write the dilaton - gravity equations (8) as the following,
\begin{eqnarray}\label{s22}
\ddot{\psi}&=&\frac{d}{2}\dot{\mu}^{2}+\frac{9-d}{2}{\dot{\nu}^2}+\frac{1}{2}\dot{\psi}^2,\nonumber\\
\ddot{\mu}&=&\dot{\mu}\dot{\psi}+\frac{1}{2}P_{rad}^{(d)}e^{\psi},\nonumber\\
\ddot{\nu}&=&\dot{\nu}\dot{\psi}\nonumber\\
\dot{\psi}^{2}&=&d\dot{\mu}^{2}+(9-d)\dot{\nu}^{2}+E_{rad}^{(d)}e^{\psi}.
\end{eqnarray}
The third equation of (22) gives,
\begin{equation}\label{s23}
\dot{\nu}=Ce^{\psi},
\end{equation}
where $C=\dot{\nu}_{0}e^{-\psi_{0}}$, in terms of initial values
$\dot{\nu}_{0}$ and $\psi_{0}$. From condition (23) we see that, if
$\dot{\nu}_{0}$ is negative then, the expansion rate for the $9-d$
small dimensions is always negative and vis versa. If
$P_{rad}^{(d)}$ in the second relation of (22) vanishes, then the
evolution of the large dimensions is similar to the small
dimensions.\\
From equations (23) it is easy to find the following equations,
\begin{eqnarray}\label{s24}
\psi&=&-\ln{[\frac{E_{rad}^{(d)}}{4}t^{2}-C_{1}t+C_{2}]},\nonumber\\
\mu&=&\frac{1}{2\sqrt{d}}\left[1-\frac{(9-d)C^{2}}{C_{1}^{2}-E_{rad}^{(d)}C_{2}}\right]^{\frac{1}{2}}
\ln{\left[\frac{\frac{E_{rad}^{(d)}}{2}t-C_{1}-\sqrt{C_{1}^{2}-E_{rad}^{(d)}C_{2}}}
{\frac{E_{rad}^{(d)}}{2}t-C_{1}+\sqrt{C_{1}^{2}-E_{rad}^{(d)}C_{2}}}\right]}+\mu_{0},\nonumber\\
\nu&=&\frac{C}{\sqrt{C_{1}^{2}-E_{rad}^{(d)}C_{2}}}\ln{\left[\frac{\frac{E_{rad}^{(d)}}{2}t-C_{1}-
\sqrt{C_{1}^{2}-E_{rad}^{(d)}C_{2}}}
{\frac{E_{rad}^{(d)}}{2}t-C_{1}+\sqrt{C_{1}^{2}-E_{rad}^{(d)}C_{2}}}\right]}+\nu_{0},
\end{eqnarray}
where
$C_{1}=\sqrt{\frac{3P_{rad}^{(d)}}{4e^{\psi_{0}}}\frac{e^{3\mu_{0}}+1}{e^{3\mu_{0}}-1}}$
and $C_{2}=e^{-\psi_{0}}$. In this case there are two scale factors
corresponding to large and small dimensions as,
\begin{eqnarray}\label{s25}
R&=&\left[\frac{\frac{E_{rad}^{(d)}}{2}t-C_{1}-\sqrt{C_{1}^{2}-E_{rad}^{(d)}C_{2}}}
{\frac{E_{rad}^{(d)}}{2}t-C_{1}+\sqrt{C_{1}^{2}-E_{rad}^{(d)}C_{2}}}\right]^{M}e^{\mu_{0}},\nonumber\\
r&=&\left[\frac{\frac{E_{rad}^{(d)}}{2}t-C_{1}-\sqrt{C_{1}^{2}-E_{rad}^{(d)}C_{2}}}
{\frac{E_{rad}^{(d)}}{2}t-C_{1}+\sqrt{C_{1}^{2}-E_{rad}^{(d)}C_{2}}}\right]^{N}e^{\nu_{0}},
\end{eqnarray}
where
$M=\frac{1}{2\sqrt{d}}\left[1-\frac{(9-d)C^{2}}{C_{1}^{2}-E_{rad}^{(d)}C_{2}}\right]^{\frac{1}{2}}$
and $N=\frac{C}{\sqrt{C_{1}^{2}-E_{rad}^{(d)}C_{2}}}$.\\\\
Also there are two Hubble expansion parameter corresponding to the
$\mu$ and $\nu$ as the following,
\begin{eqnarray}\label{s26}
H_{\mu}=\dot{\mu}&=&\sqrt{\frac{C_{1}^{2}-E_{rad}^{(d)}C_{2}-(9-d)C^{2}}{d}}
\frac{2}{E_{rad}^{(d)}t^{2}-4C_{1}t+4C_{2}},\nonumber\\
H_{\nu}=\dot{\nu}&=&\frac{4C}{E_{rad}^{(d)}t^{2}-4C_{1}t+4C_{2}}.
\end{eqnarray}
As we know our universe have three extended space dimensions, thus
it is interesting to consider $d=3$. Therefore we find four
conditions for thermal equilibrium (We take initial values of fields
as $\dot{\mu}_{0}=1$, $\mu_{0}=4$, $\dot{\nu}_{0}=-0.01$,
$\nu_{0}=0$ and $\psi_{0}=-16$ from Ref. [16]).\\
The first case is short strings in large dimensions. In this case
thermal equilibrium condition written as,
\begin{equation}\label{s27}
2(\ln
E_{rad}^{(d)})^{2}+\frac{1}{6}(E_{rad}^{(d)}-\frac{E_{rad}^{(d)}}{4}\frac{e^{3\mu_{0}}+1}{e^{3\mu_{0}}-1})e^{-\psi_{0}}
\geq-\dot{\nu}_{0}^{2}e^{-2\psi_{0}}.
\end{equation}
The second case is long strings in large dimensions. In this case
thermal equilibrium condition given by,
\begin{equation}\label{s28}
3(E_{rad}^{(d)})^{4}\frac{1+(\ln E_{rad}^{(d)})^{2}-2\ln
E_{rad}^{(d)}}{(\ln
E_{rad}^{(d)})^{2}}+(E_{rad}^{(d)}-\frac{E_{rad}^{(d)}}{4}\frac{e^{3\mu_{0}}+1}{e^{3\mu_{0}}-1})e^{-\psi_{0}}
\geq-6\dot{\nu}_{0}^{2}e^{-2\psi_{0}}.
\end{equation}
The third case is short strings in small dimensions. In this case
thermal equilibrium condition expressed as following,
\begin{equation}\label{s29}
\ln E_{rad}^{(d)}\geq\dot{\nu}_{0}^{2}e^{-2\psi_{0}}.
\end{equation}
Finally the last case is long strings in small dimensions. In this
case thermal equilibrium condition read as,
\begin{equation}\label{s30}
(E_{rad}^{(d)})^{2}(\frac{1}{2}-\frac{1}{2\ln
E_{rad}^{(d)}})\geq\dot{\nu}_{0}^{2}e^{-2\psi_{0}}.
\end{equation}
It is easy to check that, for above initial values, there is thermal
equilibrium for given any positive energy. Only condition which
break thermal equilibrium in this regime is very large negative
energy ($E\sim e^{-\psi_{0}}$).
\section{Conclusion}
In this article thermal equilibrium of the early universe in string
gas cosmology (BV scenario) dominated by closed strings
investigated. In order to find thermal equilibrium, first we
obtained closed string entropy similar to method of Ref. [14]. Then
we obtained thermal equilibrium conditions in the Hagedorn and
radiation regimes. We assumed that, range of the energy must be in
the interval $500\leq E\leq5000$ [16]. We found that in this range
there is not thermal equilibrium for both short and long strings in
small radius of Hagedorn regime. We have shown values of energy
which thermal equilibrium satisfied. We saw that for short strings
in small radius of the Hagedorn regime the total energy must be very
large to have thermal equilibrium, but for long strings in small
radius of the Hagedorn regime the small energy is sufficient to have
thermal equilibrium. On the other hand in the large radius of the
Hagedorn regime (in presence of pressure) we obtained thermal
equilibrium conditions for short and long strings. In this case we
have shown that there is possibility of avoiding Jeans instability
by using conformal time, while for cosmic time, there is Jeans
instability. Also in the case of short and long strings in the large
radius of the Hagedorn regime, by using the conformal time, we found
that there isn't thermal equilibrium for the given energy in the
range of $500\leq E\leq5000$. Finally we considered another regime
which is called radiation regime. In this regime, some of dimensions
start to expand, which denoted by $d$, and the rest remain small. We
found that there is thermal equilibrium for the case of $d=3$ in
short and long strings in large and small dimensions.

\end{document}